\documentclass[12pt]{article}
\usepackage{epsfig,amsfonts,amssymb}
\usepackage{hyperref}
\usepackage{cite}
\input epsf.sty
\usepackage{cite}

\def\ZZZ{{\hbox{ Z\kern-1.6mm Z}}}
\def\RRR{{\hbox{ R\kern-2.4mm R}}}
\def\CCC{{\hbox{ C\kern-2.0mm C}}}
\def\zzz{{\hbox{z\kern-1mm z}}}

\newcommand{\qeq}{{\hbox{=\kern-2.3mm ? \kern.5mm }}}
\renewcommand{\qeq}{=}

\newcommand{\eps}{\epsilon}

\newcommand{\GG}{{\cal G}}

\newcommand{\HH}{{\cal H}}

\newcommand{\wt}{\widetilde}
\newcommand{\wh}{\widehat}

\newcommand{\be}{\begin{equation}}
\newcommand{\ee}{\end{equation}}
\newcommand{\ben}{\begin{eqnarray}\displaystyle}
\newcommand{\een}{\end{eqnarray}}

\newcommand{\refb}[1]{(\ref{#1})}

\newcommand{\sectiono}[1]{\section{#1}\setcounter{equation}{0}}

\def\one{{\hbox{ 1\kern-.8mm l}}}
\def\zero{{\hbox{ 0\kern-1.5mm 0}}}

\newcommand{\bea}[1]{\begin{eqnarray}\label{#1} }
\newcommand{\eea}{\end{eqnarray}}

\newcommand{\eqref}{\refb}

\usepackage{bm}
\usepackage[table]{xcolor}

\def\figone{

\def\JPicScale{0.6}
\ifx\JPicScale\undefined\def\JPicScale{1}\fi
\unitlength \JPicScale mm
\begin{picture}(240,80)(0,0)
\linethickness{0.3mm}
\put(10,50){\line(1,0){100}}
\linethickness{0.3mm}
\multiput(110,50)(0.12,0.12){83}{\line(1,0){0.12}}
\linethickness{0.3mm}
\multiput(110,50)(0.12,-0.12){83}{\line(1,0){0.12}}
\linethickness{0.3mm}
\multiput(0,60)(0.12,-0.12){83}{\line(1,0){0.12}}
\linethickness{0.3mm}
\multiput(0,40)(0.12,0.12){83}{\line(1,0){0.12}}
\linethickness{0.3mm}
\put(140,50){\line(1,0){10}}
\put(170,50){\line(1,0){10}}
\put(200,50){\line(1,0){30}}
\linethickness{0.3mm}
\multiput(230,50)(0.12,0.12){83}{\line(1,0){0.12}}
\linethickness{0.3mm}
\multiput(230,50)(0.12,-0.12){83}{\line(1,0){0.12}}
\linethickness{0.3mm}
\multiput(130,60)(0.12,-0.12){83}{\line(1,0){0.12}}
\linethickness{0.3mm}
\multiput(130,40)(0.12,0.12){83}{\line(1,0){0.12}}
\linethickness{1mm}
\put(60,20){\line(0,1){60}}
\linethickness{1mm}
\put(190,20){\line(0,1){60}}
\linethickness{0.3mm}
\qbezier(20,50)(27.78,55.25)(35,55.25)
\qbezier(35,55.25)(42.22,55.25)(50,50)
\linethickness{0.3mm}
\qbezier(20,50)(27.78,44.75)(35,44.75)
\qbezier(35,44.75)(42.22,44.75)(50,50)
\linethickness{0.3mm}
\qbezier(180,50)(185.19,55.25)(190,55.25)
\qbezier(190,55.25)(194.81,55.25)(200,50)
\linethickness{0.3mm}
\qbezier(180,50)(185.19,44.75)(190,44.75)
\qbezier(190,44.75)(194.81,44.75)(200,50)
\linethickness{0.3mm}
\qbezier(150,50)(155.19,55.25)(160,55.25)
\qbezier(160,55.25)(164.81,55.25)(170,50)
\linethickness{0.3mm}
\qbezier(150,50)(155.19,44.75)(160,44.75)
\qbezier(160,44.75)(164.81,44.75)(170,50)
\put(60,10){\makebox(0,0)[cc]{(a)}}

\put(190,10){\makebox(0,0)[cc]{(b)}}

\end{picture}

}

\def\figtwo{

\def\JPicScale{0.6}
\ifx\JPicScale\undefined\def\JPicScale{1}\fi
\unitlength \JPicScale mm
\begin{picture}(150,80)(0,0)
\linethickness{0.3mm}
\put(50.68,49.75){\line(0,1){0.49}}
\multiput(50.66,50.74)(0.02,-0.49){1}{\line(0,-1){0.49}}
\multiput(50.61,51.23)(0.05,-0.49){1}{\line(0,-1){0.49}}
\multiput(50.54,51.72)(0.07,-0.49){1}{\line(0,-1){0.49}}
\multiput(50.45,52.21)(0.09,-0.49){1}{\line(0,-1){0.49}}
\multiput(50.34,52.69)(0.11,-0.48){1}{\line(0,-1){0.48}}
\multiput(50.21,53.16)(0.14,-0.47){1}{\line(0,-1){0.47}}
\multiput(50.05,53.63)(0.16,-0.47){1}{\line(0,-1){0.47}}
\multiput(49.87,54.09)(0.18,-0.46){1}{\line(0,-1){0.46}}
\multiput(49.67,54.54)(0.1,-0.23){2}{\line(0,-1){0.23}}
\multiput(49.45,54.98)(0.11,-0.22){2}{\line(0,-1){0.22}}
\multiput(49.21,55.41)(0.12,-0.22){2}{\line(0,-1){0.22}}
\multiput(48.95,55.83)(0.13,-0.21){2}{\line(0,-1){0.21}}
\multiput(48.67,56.24)(0.14,-0.2){2}{\line(0,-1){0.2}}
\multiput(48.38,56.63)(0.15,-0.2){2}{\line(0,-1){0.2}}
\multiput(48.06,57.01)(0.11,-0.13){3}{\line(0,-1){0.13}}
\multiput(47.73,57.38)(0.11,-0.12){3}{\line(0,-1){0.12}}
\multiput(47.38,57.73)(0.12,-0.12){3}{\line(1,0){0.12}}
\multiput(47.01,58.06)(0.12,-0.11){3}{\line(1,0){0.12}}
\multiput(46.63,58.38)(0.13,-0.11){3}{\line(1,0){0.13}}
\multiput(46.24,58.67)(0.2,-0.15){2}{\line(1,0){0.2}}
\multiput(45.83,58.95)(0.2,-0.14){2}{\line(1,0){0.2}}
\multiput(45.41,59.21)(0.21,-0.13){2}{\line(1,0){0.21}}
\multiput(44.98,59.45)(0.22,-0.12){2}{\line(1,0){0.22}}
\multiput(44.54,59.67)(0.22,-0.11){2}{\line(1,0){0.22}}
\multiput(44.09,59.87)(0.23,-0.1){2}{\line(1,0){0.23}}
\multiput(43.63,60.05)(0.46,-0.18){1}{\line(1,0){0.46}}
\multiput(43.16,60.21)(0.47,-0.16){1}{\line(1,0){0.47}}
\multiput(42.69,60.34)(0.47,-0.14){1}{\line(1,0){0.47}}
\multiput(42.21,60.45)(0.48,-0.11){1}{\line(1,0){0.48}}
\multiput(41.72,60.54)(0.49,-0.09){1}{\line(1,0){0.49}}
\multiput(41.23,60.61)(0.49,-0.07){1}{\line(1,0){0.49}}
\multiput(40.74,60.66)(0.49,-0.05){1}{\line(1,0){0.49}}
\multiput(40.25,60.68)(0.49,-0.02){1}{\line(1,0){0.49}}
\put(39.75,60.68){\line(1,0){0.49}}
\multiput(39.26,60.66)(0.49,0.02){1}{\line(1,0){0.49}}
\multiput(38.77,60.61)(0.49,0.05){1}{\line(1,0){0.49}}
\multiput(38.28,60.54)(0.49,0.07){1}{\line(1,0){0.49}}
\multiput(37.79,60.45)(0.49,0.09){1}{\line(1,0){0.49}}
\multiput(37.31,60.34)(0.48,0.11){1}{\line(1,0){0.48}}
\multiput(36.84,60.21)(0.47,0.14){1}{\line(1,0){0.47}}
\multiput(36.37,60.05)(0.47,0.16){1}{\line(1,0){0.47}}
\multiput(35.91,59.87)(0.46,0.18){1}{\line(1,0){0.46}}
\multiput(35.46,59.67)(0.23,0.1){2}{\line(1,0){0.23}}
\multiput(35.02,59.45)(0.22,0.11){2}{\line(1,0){0.22}}
\multiput(34.59,59.21)(0.22,0.12){2}{\line(1,0){0.22}}
\multiput(34.17,58.95)(0.21,0.13){2}{\line(1,0){0.21}}
\multiput(33.76,58.67)(0.2,0.14){2}{\line(1,0){0.2}}
\multiput(33.37,58.38)(0.2,0.15){2}{\line(1,0){0.2}}
\multiput(32.99,58.06)(0.13,0.11){3}{\line(1,0){0.13}}
\multiput(32.62,57.73)(0.12,0.11){3}{\line(1,0){0.12}}
\multiput(32.27,57.38)(0.12,0.12){3}{\line(0,1){0.12}}
\multiput(31.94,57.01)(0.11,0.12){3}{\line(0,1){0.12}}
\multiput(31.62,56.63)(0.11,0.13){3}{\line(0,1){0.13}}
\multiput(31.33,56.24)(0.15,0.2){2}{\line(0,1){0.2}}
\multiput(31.05,55.83)(0.14,0.2){2}{\line(0,1){0.2}}
\multiput(30.79,55.41)(0.13,0.21){2}{\line(0,1){0.21}}
\multiput(30.55,54.98)(0.12,0.22){2}{\line(0,1){0.22}}
\multiput(30.33,54.54)(0.11,0.22){2}{\line(0,1){0.22}}
\multiput(30.13,54.09)(0.1,0.23){2}{\line(0,1){0.23}}
\multiput(29.95,53.63)(0.18,0.46){1}{\line(0,1){0.46}}
\multiput(29.79,53.16)(0.16,0.47){1}{\line(0,1){0.47}}
\multiput(29.66,52.69)(0.14,0.47){1}{\line(0,1){0.47}}
\multiput(29.55,52.21)(0.11,0.48){1}{\line(0,1){0.48}}
\multiput(29.46,51.72)(0.09,0.49){1}{\line(0,1){0.49}}
\multiput(29.39,51.23)(0.07,0.49){1}{\line(0,1){0.49}}
\multiput(29.34,50.74)(0.05,0.49){1}{\line(0,1){0.49}}
\multiput(29.32,50.25)(0.02,0.49){1}{\line(0,1){0.49}}
\put(29.32,49.75){\line(0,1){0.49}}
\multiput(29.32,49.75)(0.02,-0.49){1}{\line(0,-1){0.49}}
\multiput(29.34,49.26)(0.05,-0.49){1}{\line(0,-1){0.49}}
\multiput(29.39,48.77)(0.07,-0.49){1}{\line(0,-1){0.49}}
\multiput(29.46,48.28)(0.09,-0.49){1}{\line(0,-1){0.49}}
\multiput(29.55,47.79)(0.11,-0.48){1}{\line(0,-1){0.48}}
\multiput(29.66,47.31)(0.14,-0.47){1}{\line(0,-1){0.47}}
\multiput(29.79,46.84)(0.16,-0.47){1}{\line(0,-1){0.47}}
\multiput(29.95,46.37)(0.18,-0.46){1}{\line(0,-1){0.46}}
\multiput(30.13,45.91)(0.1,-0.23){2}{\line(0,-1){0.23}}
\multiput(30.33,45.46)(0.11,-0.22){2}{\line(0,-1){0.22}}
\multiput(30.55,45.02)(0.12,-0.22){2}{\line(0,-1){0.22}}
\multiput(30.79,44.59)(0.13,-0.21){2}{\line(0,-1){0.21}}
\multiput(31.05,44.17)(0.14,-0.2){2}{\line(0,-1){0.2}}
\multiput(31.33,43.76)(0.15,-0.2){2}{\line(0,-1){0.2}}
\multiput(31.62,43.37)(0.11,-0.13){3}{\line(0,-1){0.13}}
\multiput(31.94,42.99)(0.11,-0.12){3}{\line(0,-1){0.12}}
\multiput(32.27,42.62)(0.12,-0.12){3}{\line(1,0){0.12}}
\multiput(32.62,42.27)(0.12,-0.11){3}{\line(1,0){0.12}}
\multiput(32.99,41.94)(0.13,-0.11){3}{\line(1,0){0.13}}
\multiput(33.37,41.62)(0.2,-0.15){2}{\line(1,0){0.2}}
\multiput(33.76,41.33)(0.2,-0.14){2}{\line(1,0){0.2}}
\multiput(34.17,41.05)(0.21,-0.13){2}{\line(1,0){0.21}}
\multiput(34.59,40.79)(0.22,-0.12){2}{\line(1,0){0.22}}
\multiput(35.02,40.55)(0.22,-0.11){2}{\line(1,0){0.22}}
\multiput(35.46,40.33)(0.23,-0.1){2}{\line(1,0){0.23}}
\multiput(35.91,40.13)(0.46,-0.18){1}{\line(1,0){0.46}}
\multiput(36.37,39.95)(0.47,-0.16){1}{\line(1,0){0.47}}
\multiput(36.84,39.79)(0.47,-0.14){1}{\line(1,0){0.47}}
\multiput(37.31,39.66)(0.48,-0.11){1}{\line(1,0){0.48}}
\multiput(37.79,39.55)(0.49,-0.09){1}{\line(1,0){0.49}}
\multiput(38.28,39.46)(0.49,-0.07){1}{\line(1,0){0.49}}
\multiput(38.77,39.39)(0.49,-0.05){1}{\line(1,0){0.49}}
\multiput(39.26,39.34)(0.49,-0.02){1}{\line(1,0){0.49}}
\put(39.75,39.32){\line(1,0){0.49}}
\multiput(40.25,39.32)(0.49,0.02){1}{\line(1,0){0.49}}
\multiput(40.74,39.34)(0.49,0.05){1}{\line(1,0){0.49}}
\multiput(41.23,39.39)(0.49,0.07){1}{\line(1,0){0.49}}
\multiput(41.72,39.46)(0.49,0.09){1}{\line(1,0){0.49}}
\multiput(42.21,39.55)(0.48,0.11){1}{\line(1,0){0.48}}
\multiput(42.69,39.66)(0.47,0.14){1}{\line(1,0){0.47}}
\multiput(43.16,39.79)(0.47,0.16){1}{\line(1,0){0.47}}
\multiput(43.63,39.95)(0.46,0.18){1}{\line(1,0){0.46}}
\multiput(44.09,40.13)(0.23,0.1){2}{\line(1,0){0.23}}
\multiput(44.54,40.33)(0.22,0.11){2}{\line(1,0){0.22}}
\multiput(44.98,40.55)(0.22,0.12){2}{\line(1,0){0.22}}
\multiput(45.41,40.79)(0.21,0.13){2}{\line(1,0){0.21}}
\multiput(45.83,41.05)(0.2,0.14){2}{\line(1,0){0.2}}
\multiput(46.24,41.33)(0.2,0.15){2}{\line(1,0){0.2}}
\multiput(46.63,41.62)(0.13,0.11){3}{\line(1,0){0.13}}
\multiput(47.01,41.94)(0.12,0.11){3}{\line(1,0){0.12}}
\multiput(47.38,42.27)(0.12,0.12){3}{\line(0,1){0.12}}
\multiput(47.73,42.62)(0.11,0.12){3}{\line(0,1){0.12}}
\multiput(48.06,42.99)(0.11,0.13){3}{\line(0,1){0.13}}
\multiput(48.38,43.37)(0.15,0.2){2}{\line(0,1){0.2}}
\multiput(48.67,43.76)(0.14,0.2){2}{\line(0,1){0.2}}
\multiput(48.95,44.17)(0.13,0.21){2}{\line(0,1){0.21}}
\multiput(49.21,44.59)(0.12,0.22){2}{\line(0,1){0.22}}
\multiput(49.45,45.02)(0.11,0.22){2}{\line(0,1){0.22}}
\multiput(49.67,45.46)(0.1,0.23){2}{\line(0,1){0.23}}
\multiput(49.87,45.91)(0.18,0.46){1}{\line(0,1){0.46}}
\multiput(50.05,46.37)(0.16,0.47){1}{\line(0,1){0.47}}
\multiput(50.21,46.84)(0.14,0.47){1}{\line(0,1){0.47}}
\multiput(50.34,47.31)(0.11,0.48){1}{\line(0,1){0.48}}
\multiput(50.45,47.79)(0.09,0.49){1}{\line(0,1){0.49}}
\multiput(50.54,48.28)(0.07,0.49){1}{\line(0,1){0.49}}
\multiput(50.61,48.77)(0.05,0.49){1}{\line(0,1){0.49}}
\multiput(50.66,49.26)(0.02,0.49){1}{\line(0,1){0.49}}

\linethickness{0.3mm}
\put(125.68,49.75){\line(0,1){0.49}}
\multiput(125.66,50.74)(0.02,-0.49){1}{\line(0,-1){0.49}}
\multiput(125.61,51.23)(0.05,-0.49){1}{\line(0,-1){0.49}}
\multiput(125.54,51.72)(0.07,-0.49){1}{\line(0,-1){0.49}}
\multiput(125.45,52.21)(0.09,-0.49){1}{\line(0,-1){0.49}}
\multiput(125.34,52.69)(0.11,-0.48){1}{\line(0,-1){0.48}}
\multiput(125.21,53.16)(0.14,-0.47){1}{\line(0,-1){0.47}}
\multiput(125.05,53.63)(0.16,-0.47){1}{\line(0,-1){0.47}}
\multiput(124.87,54.09)(0.18,-0.46){1}{\line(0,-1){0.46}}
\multiput(124.67,54.54)(0.1,-0.23){2}{\line(0,-1){0.23}}
\multiput(124.45,54.98)(0.11,-0.22){2}{\line(0,-1){0.22}}
\multiput(124.21,55.41)(0.12,-0.22){2}{\line(0,-1){0.22}}
\multiput(123.95,55.83)(0.13,-0.21){2}{\line(0,-1){0.21}}
\multiput(123.67,56.24)(0.14,-0.2){2}{\line(0,-1){0.2}}
\multiput(123.38,56.63)(0.15,-0.2){2}{\line(0,-1){0.2}}
\multiput(123.06,57.01)(0.11,-0.13){3}{\line(0,-1){0.13}}
\multiput(122.73,57.38)(0.11,-0.12){3}{\line(0,-1){0.12}}
\multiput(122.38,57.73)(0.12,-0.12){3}{\line(1,0){0.12}}
\multiput(122.01,58.06)(0.12,-0.11){3}{\line(1,0){0.12}}
\multiput(121.63,58.38)(0.13,-0.11){3}{\line(1,0){0.13}}
\multiput(121.24,58.67)(0.2,-0.15){2}{\line(1,0){0.2}}
\multiput(120.83,58.95)(0.2,-0.14){2}{\line(1,0){0.2}}
\multiput(120.41,59.21)(0.21,-0.13){2}{\line(1,0){0.21}}
\multiput(119.98,59.45)(0.22,-0.12){2}{\line(1,0){0.22}}
\multiput(119.54,59.67)(0.22,-0.11){2}{\line(1,0){0.22}}
\multiput(119.09,59.87)(0.23,-0.1){2}{\line(1,0){0.23}}
\multiput(118.63,60.05)(0.46,-0.18){1}{\line(1,0){0.46}}
\multiput(118.16,60.21)(0.47,-0.16){1}{\line(1,0){0.47}}
\multiput(117.69,60.34)(0.47,-0.14){1}{\line(1,0){0.47}}
\multiput(117.21,60.45)(0.48,-0.11){1}{\line(1,0){0.48}}
\multiput(116.72,60.54)(0.49,-0.09){1}{\line(1,0){0.49}}
\multiput(116.23,60.61)(0.49,-0.07){1}{\line(1,0){0.49}}
\multiput(115.74,60.66)(0.49,-0.05){1}{\line(1,0){0.49}}
\multiput(115.25,60.68)(0.49,-0.02){1}{\line(1,0){0.49}}
\put(114.75,60.68){\line(1,0){0.49}}
\multiput(114.26,60.66)(0.49,0.02){1}{\line(1,0){0.49}}
\multiput(113.77,60.61)(0.49,0.05){1}{\line(1,0){0.49}}
\multiput(113.28,60.54)(0.49,0.07){1}{\line(1,0){0.49}}
\multiput(112.79,60.45)(0.49,0.09){1}{\line(1,0){0.49}}
\multiput(112.31,60.34)(0.48,0.11){1}{\line(1,0){0.48}}
\multiput(111.84,60.21)(0.47,0.14){1}{\line(1,0){0.47}}
\multiput(111.37,60.05)(0.47,0.16){1}{\line(1,0){0.47}}
\multiput(110.91,59.87)(0.46,0.18){1}{\line(1,0){0.46}}
\multiput(110.46,59.67)(0.23,0.1){2}{\line(1,0){0.23}}
\multiput(110.02,59.45)(0.22,0.11){2}{\line(1,0){0.22}}
\multiput(109.59,59.21)(0.22,0.12){2}{\line(1,0){0.22}}
\multiput(109.17,58.95)(0.21,0.13){2}{\line(1,0){0.21}}
\multiput(108.76,58.67)(0.2,0.14){2}{\line(1,0){0.2}}
\multiput(108.37,58.38)(0.2,0.15){2}{\line(1,0){0.2}}
\multiput(107.99,58.06)(0.13,0.11){3}{\line(1,0){0.13}}
\multiput(107.62,57.73)(0.12,0.11){3}{\line(1,0){0.12}}
\multiput(107.27,57.38)(0.12,0.12){3}{\line(1,0){0.12}}
\multiput(106.94,57.01)(0.11,0.12){3}{\line(0,1){0.12}}
\multiput(106.62,56.63)(0.11,0.13){3}{\line(0,1){0.13}}
\multiput(106.33,56.24)(0.15,0.2){2}{\line(0,1){0.2}}
\multiput(106.05,55.83)(0.14,0.2){2}{\line(0,1){0.2}}
\multiput(105.79,55.41)(0.13,0.21){2}{\line(0,1){0.21}}
\multiput(105.55,54.98)(0.12,0.22){2}{\line(0,1){0.22}}
\multiput(105.33,54.54)(0.11,0.22){2}{\line(0,1){0.22}}
\multiput(105.13,54.09)(0.1,0.23){2}{\line(0,1){0.23}}
\multiput(104.95,53.63)(0.18,0.46){1}{\line(0,1){0.46}}
\multiput(104.79,53.16)(0.16,0.47){1}{\line(0,1){0.47}}
\multiput(104.66,52.69)(0.14,0.47){1}{\line(0,1){0.47}}
\multiput(104.55,52.21)(0.11,0.48){1}{\line(0,1){0.48}}
\multiput(104.46,51.72)(0.09,0.49){1}{\line(0,1){0.49}}
\multiput(104.39,51.23)(0.07,0.49){1}{\line(0,1){0.49}}
\multiput(104.34,50.74)(0.05,0.49){1}{\line(0,1){0.49}}
\multiput(104.32,50.25)(0.02,0.49){1}{\line(0,1){0.49}}
\put(104.32,49.75){\line(0,1){0.49}}
\multiput(104.32,49.75)(0.02,-0.49){1}{\line(0,-1){0.49}}
\multiput(104.34,49.26)(0.05,-0.49){1}{\line(0,-1){0.49}}
\multiput(104.39,48.77)(0.07,-0.49){1}{\line(0,-1){0.49}}
\multiput(104.46,48.28)(0.09,-0.49){1}{\line(0,-1){0.49}}
\multiput(104.55,47.79)(0.11,-0.48){1}{\line(0,-1){0.48}}
\multiput(104.66,47.31)(0.14,-0.47){1}{\line(0,-1){0.47}}
\multiput(104.79,46.84)(0.16,-0.47){1}{\line(0,-1){0.47}}
\multiput(104.95,46.37)(0.18,-0.46){1}{\line(0,-1){0.46}}
\multiput(105.13,45.91)(0.1,-0.23){2}{\line(0,-1){0.23}}
\multiput(105.33,45.46)(0.11,-0.22){2}{\line(0,-1){0.22}}
\multiput(105.55,45.02)(0.12,-0.22){2}{\line(0,-1){0.22}}
\multiput(105.79,44.59)(0.13,-0.21){2}{\line(0,-1){0.21}}
\multiput(106.05,44.17)(0.14,-0.2){2}{\line(0,-1){0.2}}
\multiput(106.33,43.76)(0.15,-0.2){2}{\line(0,-1){0.2}}
\multiput(106.62,43.37)(0.11,-0.13){3}{\line(0,-1){0.13}}
\multiput(106.94,42.99)(0.11,-0.12){3}{\line(0,-1){0.12}}
\multiput(107.27,42.62)(0.12,-0.12){3}{\line(1,0){0.12}}
\multiput(107.62,42.27)(0.12,-0.11){3}{\line(1,0){0.12}}
\multiput(107.99,41.94)(0.13,-0.11){3}{\line(1,0){0.13}}
\multiput(108.37,41.62)(0.2,-0.15){2}{\line(1,0){0.2}}
\multiput(108.76,41.33)(0.2,-0.14){2}{\line(1,0){0.2}}
\multiput(109.17,41.05)(0.21,-0.13){2}{\line(1,0){0.21}}
\multiput(109.59,40.79)(0.22,-0.12){2}{\line(1,0){0.22}}
\multiput(110.02,40.55)(0.22,-0.11){2}{\line(1,0){0.22}}
\multiput(110.46,40.33)(0.23,-0.1){2}{\line(1,0){0.23}}
\multiput(110.91,40.13)(0.46,-0.18){1}{\line(1,0){0.46}}
\multiput(111.37,39.95)(0.47,-0.16){1}{\line(1,0){0.47}}
\multiput(111.84,39.79)(0.47,-0.14){1}{\line(1,0){0.47}}
\multiput(112.31,39.66)(0.48,-0.11){1}{\line(1,0){0.48}}
\multiput(112.79,39.55)(0.49,-0.09){1}{\line(1,0){0.49}}
\multiput(113.28,39.46)(0.49,-0.07){1}{\line(1,0){0.49}}
\multiput(113.77,39.39)(0.49,-0.05){1}{\line(1,0){0.49}}
\multiput(114.26,39.34)(0.49,-0.02){1}{\line(1,0){0.49}}
\put(114.75,39.32){\line(1,0){0.49}}
\multiput(115.25,39.32)(0.49,0.02){1}{\line(1,0){0.49}}
\multiput(115.74,39.34)(0.49,0.05){1}{\line(1,0){0.49}}
\multiput(116.23,39.39)(0.49,0.07){1}{\line(1,0){0.49}}
\multiput(116.72,39.46)(0.49,0.09){1}{\line(1,0){0.49}}
\multiput(117.21,39.55)(0.48,0.11){1}{\line(1,0){0.48}}
\multiput(117.69,39.66)(0.47,0.14){1}{\line(1,0){0.47}}
\multiput(118.16,39.79)(0.47,0.16){1}{\line(1,0){0.47}}
\multiput(118.63,39.95)(0.46,0.18){1}{\line(1,0){0.46}}
\multiput(119.09,40.13)(0.23,0.1){2}{\line(1,0){0.23}}
\multiput(119.54,40.33)(0.22,0.11){2}{\line(1,0){0.22}}
\multiput(119.98,40.55)(0.22,0.12){2}{\line(1,0){0.22}}
\multiput(120.41,40.79)(0.21,0.13){2}{\line(1,0){0.21}}
\multiput(120.83,41.05)(0.2,0.14){2}{\line(1,0){0.2}}
\multiput(121.24,41.33)(0.2,0.15){2}{\line(1,0){0.2}}
\multiput(121.63,41.62)(0.13,0.11){3}{\line(1,0){0.13}}
\multiput(122.01,41.94)(0.12,0.11){3}{\line(1,0){0.12}}
\multiput(122.38,42.27)(0.12,0.12){3}{\line(0,1){0.12}}
\multiput(122.73,42.62)(0.11,0.12){3}{\line(0,1){0.12}}
\multiput(123.06,42.99)(0.11,0.13){3}{\line(0,1){0.13}}
\multiput(123.38,43.37)(0.15,0.2){2}{\line(0,1){0.2}}
\multiput(123.67,43.76)(0.14,0.2){2}{\line(0,1){0.2}}
\multiput(123.95,44.17)(0.13,0.21){2}{\line(0,1){0.21}}
\multiput(124.21,44.59)(0.12,0.22){2}{\line(0,1){0.22}}
\multiput(124.45,45.02)(0.11,0.22){2}{\line(0,1){0.22}}
\multiput(124.67,45.46)(0.1,0.23){2}{\line(0,1){0.23}}
\multiput(124.87,45.91)(0.18,0.46){1}{\line(0,1){0.46}}
\multiput(125.05,46.37)(0.16,0.47){1}{\line(0,1){0.47}}
\multiput(125.21,46.84)(0.14,0.47){1}{\line(0,1){0.47}}
\multiput(125.34,47.31)(0.11,0.48){1}{\line(0,1){0.48}}
\multiput(125.45,47.79)(0.09,0.49){1}{\line(0,1){0.49}}
\multiput(125.54,48.28)(0.07,0.49){1}{\line(0,1){0.49}}
\multiput(125.61,48.77)(0.05,0.49){1}{\line(0,1){0.49}}
\multiput(125.66,49.26)(0.02,0.49){1}{\line(0,1){0.49}}

\linethickness{0.3mm}
\qbezier(50,55)(65.58,62.88)(78.81,62.88)
\qbezier(78.81,62.88)(92.05,62.88)(105,55)
\linethickness{0.3mm}
\linethickness{0.3mm}
\qbezier(50,45)(65.58,37.12)(78.81,37.12)
\qbezier(78.81,37.12)(92.05,37.12)(105,45)
\linethickness{1mm}
\put(80,20){\line(0,1){60}}
\put(40,50){\makebox(0,0)[cc]{$\Gamma^{(N_1)}$}}

\put(115,50){\makebox(0,0)[cc]{$\Gamma^{(N_2)}$}}

\put(70,58){\makebox(0,0)[cc]{$\bullet$}}

\put(70,50){\makebox(0,0)[cc]{$\bullet$}}

\put(70,42){\makebox(0,0)[cc]{$\bullet$}}

\put(20,54){\makebox(0,0)[cc]{$\bullet$}}

\put(20,48){\makebox(0,0)[cc]{$\bullet$}}

\put(20,40){\makebox(0,0)[cc]{$\bullet$}}

\put(135,58){\makebox(0,0)[cc]{$\bullet$}}

\put(135,50){\makebox(0,0)[cc]{$\bullet$}}

\put(135,42){\makebox(0,0)[cc]{$\bullet$}}

\linethickness{0.3mm}
\multiput(10,65)(0.24,-0.12){83}{\line(1,0){0.24}}
\linethickness{0.3mm}
\linethickness{0.3mm}
\multiput(10,30)(0.3,0.12){84}{\line(1,0){0.3}}
\linethickness{0.3mm}
\multiput(120,60)(0.36,0.12){83}{\line(1,0){0.36}}
\linethickness{0.3mm}
\multiput(120,40)(0.36,-0.12){83}{\line(1,0){0.36}}
\end{picture}

}

\def\figblind{

\def\JPicScale{0.6}
\ifx\JPicScale\undefined\def\JPicScale{1}\fi
\unitlength \JPicScale mm

}

\def\figpmod{

\def\JPicScale{0.6}
\ifx\JPicScale\undefined\def\JPicScale{1}\fi
\unitlength \JPicScale mm

}

\begin{document}

\baselineskip 24pt

\begin{center}
{\Large \bf  Unitarity of Superstring  Field Theory}

\end{center}

\vskip .6cm
\medskip

\vspace*{4.0ex}

\baselineskip=18pt

\centerline{\large \rm Ashoke Sen}

\vspace*{4.0ex}

\centerline{\large \it Harish-Chandra Research Institute}
\centerline{\large \it  Chhatnag Road, Jhusi,
Allahabad 211019, India}

\centerline{and}

\centerline{\large \it Homi Bhabha National Institute}
\centerline{\large \it Training School Complex, Anushakti Nagar,
    Mumbai 400085, India}

\vspace*{1.0ex}
\centerline{\small E-mail:  sen@mri.ernet.in}

\vspace*{5.0ex}

\centerline{\bf Abstract} \bigskip

We complete the proof of unitarity of (compactified) 
heterotic and type II string field theories by showing that in the cut diagrams
only physical states appear in the sum over intermediate states. This analysis
takes into account the effect of mass and wave-function renormalization, and
the possibility that the true vacuum may be related to the perturbative vacuum by
small shifts in the string fields.

\vfill \eject

\baselineskip 18pt

\tableofcontents

\sectiono{Introduction} \label{sintro}

In a previous paper \cite{1604.01783} 
we derived the Cutkosky rules for superstring field theory under
the assumption that the action for string field theory is real. This assumption was 
proved later\cite{1606.03455}. 
Cutkosky rules derived in \cite{1604.01783} 
establish that the T-matrix -- related to the S-matrix via the relation 
$S=1-iT$ -- satisfies the relation $i(T-T^\dagger) = T^\dagger
|n\rangle\langle n|T$, where the sum over $|n\rangle$ runs over all states in the
Siegel gauge. This would establish unitarity of the theory 
if all states in the Siegel gauge were physical states, 
However since string field theory is a gauge theory, Cutkosky rules
do not automatically prove unitarity. A cut propagator representing
$|n\rangle \langle n|$, 
besides propagating physical on-shell
intermediate states, also has unphysical and pure gauge states. Therefore in order
to prove unitarity we need to prove that the sum over intermediate states in a cut 
diagram receives contribution from only the physical on-shell states, and the contribution
from all other states cancel. This is what we shall show in this paper.

Earlier attempts\cite{giddings,dhoker} to prove unitarity of superstring theory
in the covariant formulation 
relied on proving equivalence to light-cone string field 
theory\cite{mandelstam1,mandelstam2}. However since
light-cone superstring field theory encounters contact term 
divergences\cite{gr1,gr2,gr3,greenseiberg}, it is not
clear if this can be lifted to a valid proof after 
taking into account the various subtleties of the covariant formulation described
in \cite{1209.5461}. 
Some recent attempts to circumvent this
difficulty can be found in \cite{1605.04666}. 
Another systematic procedure for computing the
imaginary part of the string theory amplitude is the $i\eps$ prescription of 
\cite{berera,1307.5124},
but it is not clear at this stage
how this can be used to prove unitarity of the amplitude.

The rest of the paper is organized as follows. In \S\ref{sprop} we derive some useful
properties of the quantum corrected propagator of string field theory and residues at
its poles. In \S\ref{sproof} we use the Ward identities of string field theory to show that
only physical states contribute in the sum over intermediate states in a cut diagram. 
In \S\ref{sdiss} we discuss some  open problems.

\sectiono{The propagator} \label{sprop}

The main tool in our analysis
will be the full propagator computed from the
one particle irreducible (1PI) effective
action of superstring field theory.
We shall follow the conventions of \cite{1508.02481}, and 
begin by collecting some basic results in 1PI effective string field theory as
reviewed in \cite{1508.02481}. This will be followed by a review of
some basic results in
Cutkosky rules derived in \cite{1604.01783}.. Finally we shall combine these
results to derive the general form of the contribution from a cut propagator 
as given in \refb{edel0}.

We shall work with the heterotic string theory for
simplicity and will describe the generalization to type II string theories at the end
of the section.
We define
$\HH_T$ 
to be the subspace of GSO even states in the matter ghost conformal field theory
satisfying
\ben
&& b_0^- |s\rangle =0, \quad L_0^- |s\rangle=0, \quad \hbox{for $|s\rangle\in \HH_T$}
\, . 
\een
Here $b_n,\bar b_n,c_n, \bar c_n$ are the modes of the usual $b,\bar b,c,\bar c$ ghost
fields, $L_n,\bar L_n$ are the total Virasoro generators, and
\be
b_0^\pm =b_0\pm \bar b_0, \quad c_0^\pm ={1\over 2} (c_0\pm \bar c_0), \quad
L_0^\pm = L_0\pm \bar L_0\, .
\ee
$\HH_n$ will denote the subspace
of states in $\HH_T$ carrying picture number $n$. 

Even though we follow closely the formalism described in \cite{1508.02481}, 
there are two ways in which the action that we consider differs from the one
analyzed in \cite{1508.02481}:
\begin{enumerate}
\item We shall implicitly assume that the 1PI effective action
we use 
comes from the sum of 1PI diagrams  of the action described in \cite{1508.05387} so that the
Cutkosky rules hold\cite{1604.01783}.
This means that the class of actions we shall consider will be more restrictive than the
ones used in \cite{1508.02481}. But this does not prevent us from using the results
derived in \cite{1508.02481} since the latter describes a more general class of
theories.
\item
In the analysis of
\cite{1508.02481} we had restricted the interacting 
string field to carry ghost number 2 and picture numbers 
$-1$ or $-1/2$.\footnote{The formulation of the theory 
given in \cite{1508.05387} also requires us
to introduce a free string field taking value in $\HH_{-1}\oplus\HH_{-3/2}$, 
but this will not play any
role in our analysis.} 
Here
we shall keep the form of the action unchanged but 
allow the string field to carry all possible ghost numbers since
we want to include in our analysis not only the matter fields but also the ghost fields
as external states.\footnote{The corresponding action will describe the `1PI master action' 
and
will satisfy the classical master equation like the classical master 
action\cite{9206084,1508.05387}. As in \cite{9206084,1508.05387}, 
the master action
is obtained from the original action by relaxing the constraint on ghost number of the
string field, but keeping the form of the action the same.}
It is easy to verify that the results of \cite{1508.02481} that we shall be using, namely
eq.\refb{edefDelta} below
for the propagator and eq.\refb{egrid} below for the truncated Green's function, 
are valid for these more general string fields.
\end{enumerate}

The kinetic operator $\wh Q_B$ of 1PI effective string field theory around
the quantum corrected vacuum, and a related
operator $\wt Q_B$ introduced in \cite{1508.02481} will play special roles
in our analysis. 
$\wh Q_B$ and $\wt Q_B$ are operators of ghost number 1, 
acting respectively on the states in 
\be \label{eht}
\wh\HH_T\equiv \HH_{-1}\oplus \HH_{-1/2} \quad \hbox{and} \quad 
\wt\HH_T\equiv \HH_{-1}\oplus \HH_{-3/2}\, ,
\ee
producing
states in $\wh\HH_T$ and $\wt\HH_T$ respectively. 
$\wh Q_B$ and $\wt Q_B$  have the form
\be \label{eqbk}
\wh Q_B = Q_B + \GG \, K, \quad \wt Q_B = Q_B + K \, \GG\, .
\ee
Here $Q_B$ is the nilpotent BRST operator. 
$K$ is some operator  that acts on states in $\wh \HH_T$ and produces
states in $\wt \HH_T$. It is related to the 1PI two point function and
can be computed using perturbation theory. 
$\GG$ is the identity operator in the Neveu-Schwarz (NS) sector and the 
zero mode
of the picture changing operator (PCO) in the Ramond (R) sector, and satisfies
\be
[\GG, b_0^\pm]=0\, , \qquad [\GG, L_0^\pm]=0, \qquad [\GG, Q_B]=0\, .
\ee
$\wh Q_B$ and $\wt Q_B$ satisfy
\be \label{enil}
\wh Q_B^2=0, \quad \wt Q_B^2=0\, ,
\ee
and
\ben \label{eabqb}
&& \langle A| c_0^- \wh Q_B = (-1)^{\gamma_A} 
\langle \wt Q_B A| c_0^-  \quad \hbox{for $|A\rangle\in \wt\HH_T$}  \, , \nonumber \\
&& \langle B| c_0^- \wt Q_B = (-1)^{\gamma_B} 
\langle \wh Q_B B| c_0^-  
  \quad \hbox{for $|B\rangle\in \wh\HH_T$}\, ,
\een
where $\gamma_A$ and $\gamma_B$ are grassmannalities of $A$ and $B$, and 
$\langle \wt Q_B A|$ and $\langle \wh Q_B B|$ are  respectively
the BPZ conjugates of
$\wt Q_B|A\rangle$ and $\wh Q_B|B\rangle$.
It follows from \refb{eqbk} that $\wh Q_B \GG = \GG \wt Q_B$.

While the interacting string field takes value in $\wh\HH_T$, Siegel gauge condition
further restricts the string field to be annihilated by $b_0^+$. 
The full Siegel gauge propagator $\Delta$ for the interacting string field of the
1PI effective action was constructed in \cite{1508.02481}. 
It acts on states in $c_0^- \wt\HH_T$ and  produces states in $\wh\HH_T$, and
has the form
\be \label{edefDelta}
\Delta = \beta\, \GG (L_0^+ +  b_0^+ K\GG)^{-1} b_0^+ b_0^- 
= \beta\, \GG \, b_0^+   (L_0^+ +  K\GG \, b_0^+ )^{-1}  b_0^-  
\quad \hbox{
acting on states in $c_0^-\wt\HH_T$}\, .
\ee
Here $\beta$ is a constant that depends on the normalization of the action and
includes a factor of $i$ for Lorentzian signature space-time background.
$\Delta$ satisfies
\be \label{eb0pm}
b_0^+ \Delta = 0, \quad \Delta \, b_0^+ = 0\, ,
\ee
\be \label{eboth}
\wh Q_B \Delta c_0^-+ \Delta c_0^- \wt Q_B =  \beta\, \GG \quad \hbox{
acting on states in $\wt\HH_T$}\, .
\ee

Construction of $\Delta$ requires inverting the operator
$L_0^+ +  b_0^+ K\GG$ 
which is an infinite dimensional matrix. 
At generic momentum we can evaluate $\Delta$ using perturbation theory in $K$ 
leading to
a sum over Feynman diagrams contributing to the off-shell two point function.
However when the momentum is close to a value where $L_0^+$ vanishes for some
states, perturbation theory breaks down since diagrams of higher order will
have poles of higher order. In this case a
systematic procedure for computing $\Delta$  
was described in \cite{1401.7014}. 
The strategy is to 
work at some fixed mass$^2$ level $L$ (defined by the 
momentum independent contribution to $L_0^+$)
and `integrate out' the contribution to $\Delta$ from
fields in the other mass levels. The latter operation 
can be carried out perturbatively. 
This generates a finite dimensional matrix in the space
of states at the mass$^2$ level $L$ which can then be inverted explicitly.
More explicitly, this corresponds to using the identity
\be \label{ealgo}
\pmatrix{A & B\cr C & D}^{-1} =\pmatrix{ (A - B D^{-1} C)^{-1} & 
-(A - B D^{-1} C )^{-1} B D^{-1}
\cr
- D^{-1} C (A - B D^{-1} C)^{-1} & D^{-1} + D^{-1} C (A - B D^{-1} C)^{-1} B D^{-1}}
\ee
where $A$ and $D$ are square matrices and $B$ and $C$ are rectangular matrices.
In our case $A$ denotes the kinetic operator at mass$^2$ level $L$ and is a finite
dimensional matrix, $D$ is the kinetic operator for all mass$^2$ levels other than $L$
and $B$ and $C$ are the mixing matrices between mass$^2$ level $L$ and mass$^2$ 
level
other than $L$. $D^{-1}$ can be computed perturbatively to any given order, and 
we compute $(A - B D^{-1} C )^{-1}$ by exact matrix inversion.
This matrix then can be used {\it e.g.}  to find the  poles of $\Delta$ in the $k^2$ 
plane for $-k^2$ near mass$^2$ level $L$.

\begin{figure}
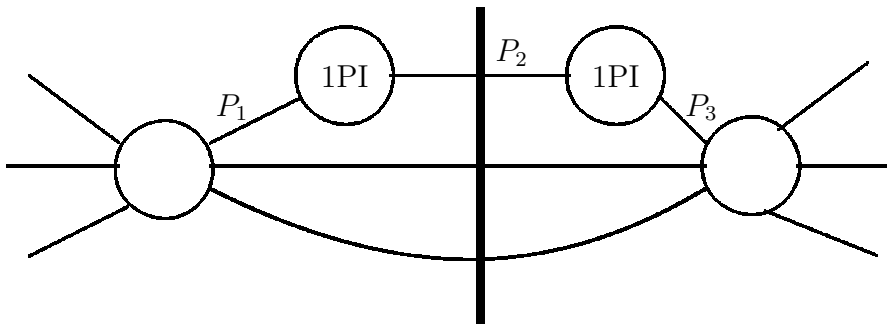


\begin{center}

\figblind

\end{center}

\vskip -1in

\caption{A problematic cut diagram.
\label{fonemod}}
\end{figure}

We now briefly review the results of \cite{1604.01783}.
The analysis of \cite{1604.01783}  tells us that the string field theory amplitudes obey
the Cutkosky cutting rules. 
These rules may be summarized as follows.\footnote{Here we are considering the
Feynman diagrams of superstring field theory expanded around the shifted
background, and not that of 1PI effective string
field theory. For this reason the vertices are hermitian, we need to include loop diagrams,
and Cutkosky rules hold. If we use the 1PI effective action to compute the amplitude,
then there are no loop diagrams,
but the verices will be complex and part of the contribution to the ant--hermitian part
of an amplitude will arise from the anti-hermitian part of the vertices.}
If we draw a  Feynman diagram with the incoming states on the
left and the outgoing states on the right, then the contribution to $i(T-T^\dagger)$ is
given by the sum over all cuts of the diagram where a cut is a line through the
diagram separating the incoming states from the outgoing states. The rule for 
computing a cut diagram is to replace a cut internal propagator $- i (k^2+m_0^2)^{-1}$
by $2\pi \delta (k^2+m_0^2) \theta(k^0)$. Here 
$m_0$ is the tree level mass,
$k$ is the momentum flowing through the cut propagator from the left to the 
right and $\theta$ denotes step function. Furthermore,
the amplitude to the right of the cut is hermitian conjugated. 
However, naive application of this result will give
divergent result from cut diagrams of the form shown in Fig.~\ref{fonemod}.  The cut passing
through the propagator $P_2$ forces the momentum passing through this to be on-shell,
but this also forces the momentum passing through the uncut propagators $P_1$ and $P_3$
to be on-shell, making them diverge. The remedy suggested in \cite{veltman,diagrammar}
is to sum over all cuts of a propagator to express the result as the hermitian part of the
full propagator. It is simplest to illustrate this through a scalar field propagator. Let us
suppose that
the full quantum corrected propagator has the form

\be 
\Delta = -i\, (k^2+m_0^2+\Gamma(k)-i\eps)^{-1} \, ,
\ee
where $m_0$ is the tree level mass, $k$ is the momentum flowing from the 
left to the right,
and $-i\Gamma(k)$ represents the contribution
from the 1PI two point function. Then the sum over all cuts of the propagator 
can be expressed as 
\be \label{epropim}
-i \, \left[ (k^2+m_0^2+\Gamma(k)-i\eps)^{-1} -  (k^2+m_0^2+\Gamma(k)^*+i\eps)^{-1}
\right]\, \theta(k^0)\, .
\ee
There are three cases to be considered. If for the value of $k$ of interest $\Gamma(k)$ has
an imaginary part, then we can ignore the $i\eps$ term and express \refb{epropim} as
\be \label{epp1}
-i (k^2+m_0^2+\Gamma(k))^{-1}\,  i (\Gamma(k) -\Gamma(k)^*) \, \theta(k^0)\, 
\, i (k^2+m_0^2+\Gamma(k)^*)^{-1}\, .
\ee
If $\Gamma(k)$ is
real and $k^2+m_0^2+\Gamma(k)$ is away from 0, then \refb{epropim} vanishes.
Finally if $\Gamma(k)$ is real and the full propagator has 
a pole on the real $k^2$ axis at $k^2+m^2=0$
with residue $-i\, Z$, then \refb{epropim} behaves as
\be \label{epp2}
2\pi \, Z\, \delta(k^2+m^2) \, \theta(k^0)\, ,
\ee
near $k^2+m^2=0$.  Therefore \refb{epropim} may be expressed as the sum
of \refb{epp1} and \refb{epp2}. This can be represented diagrammatically as
in Fig.~\ref{fpmod}.

\begin{figure}

\begin{center}

\hbox{\figpmod}

\end{center}

\vskip -1in

\caption{Diagrammatic representation of \refb{epropim}-\refb{epp2}. The left
hand side represents \refb{epropim}, the first term on the right hand side represents
\refb{epp1} and the second term on the right hand side represents \refb{epp2}.
\label{fpmod}}
\end{figure}

For the full superstring field theory this has the following consequence. Let us suppose
that the full propagator $\Delta$ has
a pole at $k^2+m^2=0$. Then near $k^2=-m^2$ we have
\be 
\Delta = -i\, (k^2+m^2-i\eps)^{-1} \Delta_0 + \hbox{non-singular}\, .
\ee
The $i\eps$ determines the side of the integration contour on which the 
pole lies\cite{1604.01783}.
Even though
$\Delta$ is an infinite dimensional matrix, $\Delta_0$ is a matrix of finite rank
since for given momentum we expect only a finite number of states 
for which the propagator develops a pole at $k^2=-m^2$.
The rules for computing the contribution from a cut propagator
can be summarized as follows:

\begin{figure}
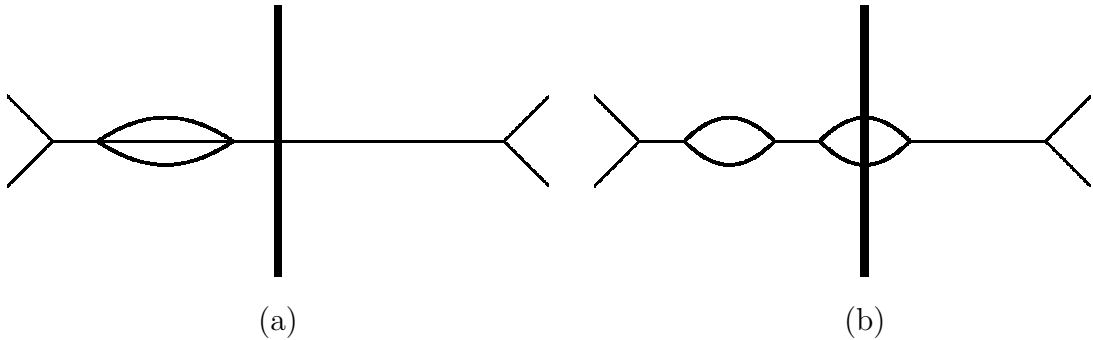


\begin{center}

\figone

\end{center}

\vskip -.2in

\caption{Fig.~(a) shows the example of a disallowed cut diagram and Fig.~(b) shows the
example of an allowed cut diagram. In both examples the thick vertical line denotes the 
cut.
\label{f1}}
\end{figure}

\begin{enumerate}
\item If $m^2$ is real then the corresponding cut propagator should be replaced by
\be\label{epole}
2\pi \, \delta(k^2+m^2) \, \theta(k^0) \, \Delta_0\, ,
\ee
where $k$ is the momentum carried by the 
cut propagator from the left side of the cut to the right side.
Furthermore we do not include any virtual self-energy corrections on either side of
a cut propagator, {\it e.g.} the diagrams of the type 
shown in Fig.~\ref{f1}(a) are not allowed
since their contribution has already been
included in the $\Delta_0$ factor in \refb{epole} and in the use of the
renormalized mass
$m$. 
However a cut can pass through a self energy diagram, {\it e.g.} a diagram of the
form shown in Fig.~\ref{f1}(b) is allowed. These diagrams capture the imaginary part
of the propagator other than the pole contribution \refb{epole}, as represented by
\refb{epp1}.
\item If $m^2$ has an imaginary part, i.e.\ if the pole corresponds to an unstable
particle, then we do not 
have the contribution \refb{epole}. This is consistent with the fact that unstable
particles are not genuine asymptotic states and should not appear in the sum over
intermediate states in a unitarity relation.
However cuts passing through the self energy
diagrams, like the ones shown in Fig.~\ref{f1}(b) are still allowed.
\item  \label{i3}
If the momentum $k$ carried by a
virtual uncut 
propagator is near a pole of the propagator, 
we must use the resummed
propagator that includes repeated insertion of one particle irreducible (1PI)
self-energy diagrams on the propagator. For 
our analysis this means that near $k^2+m^2=0$
we should use the propagator $\Delta$ 
constructed using \refb{edefDelta}, \refb{ealgo} which already has
resummation built into it.\footnote{This rule is particularly important for unstable 
particles as can be illustrated using the example of Fig.~\ref{f1}(b). If
the intermediate single particle state represented by the
horizontal line corresponds to an unstable particle, it is kinematically possible
for momentum flowing through the horizontal 
line to be near its classical on-shell value, and
repeated insertion of self-energy diagrams on this will generate divergences
of arbitrarily high order. Therefore we must use the resummed propagator for
which we only have first order pole and the
pole is shifted away from the real axis. In the limit of zero string coupling, the combined
contribution from the resummed propagators on two sides of the cut and the
contribution from the cut 1PI two point function approaches the delta function
contribution given in \refb{epole}.}
\end{enumerate}

Therefore for analyzing cut diagrams
we need to focus on the properties of $\Delta_0$ associated
with the poles that occur at real momenta. 
Multiplying both sides of \refb{eb0pm}, \refb{eboth} by 
$k^2+m^2$ and taking the limit $k^2\to -m^2$, we get
\be \label{eb0pma}
b_0^+ \Delta_0=0, \quad \Delta_0 b_0^+=0\, ,
\ee
\be \label{eqdid}
\wh Q_B \Delta_0 c_0^-+ \Delta_0 c_0^- \wt Q_B = 0\, .
\ee
Multiplying \refb{eqdid} by $b_0^+$ from left/right and using \refb{eb0pma} we get
\be \label{eid1}
b_0^+ \wh Q_B  \Delta_0 c_0^- =0, \quad
\Delta_0  c_0^- \wt Q_B b_0^+ = 0\, .
\ee
Let us now use a general ansatz
\be
\Delta_0=\sum_{m=1}^R |\Phi_m\rangle \langle \Psi_m| 
\ee
where $R$ is the rank of $\Delta_0$ and $\{|\Phi_m\rangle\}$ and $\{\langle\Psi_m|\}$
are a set of linearly independent states. Since $\Delta_0$ acts on states in 
$c_0^-\wt\HH_T$ to produce states in $\wh\HH_T$, and BPZ inner product pairs states
in $\wh\HH_T$ with states in $c_0^-\wt\HH_T$, we have
\be
|\Phi_m\rangle\in\wh\HH_T, \qquad |\Psi_m\rangle\in\wh\HH_T\, .
\ee
Eqs.~\refb{eb0pma}, \refb{eid1}
now give
\be \label{epsbm}
b_0^+ |\Phi_m\rangle =0,  \quad \langle \Psi_m| b_0^+=0, \quad
\quad b_0^+ \wh Q_B |\Phi_m\rangle = 0, \quad \langle \Psi_m| c_0^- \wt Q_B b_0^+=0
\Rightarrow \langle  \wh Q_B\Psi_m| b_0^+=0\, ,
\ee
where in the last step we have used \refb{eabqb} and the fact that 
$b_0^- \wh Q_B|\Psi_m\rangle=0$.
\refb{epsbm} is just a reflection of the fact that the poles of $\Delta$ are associated
with zero eigenvalues of the kinetic operator in the Siegel gauge.

We now classify the candidates for $|\Phi_m\rangle$
satisfying these conditions near a particular pole.
\begin{enumerate}
\item Unphysical states: These are linearly independent 
states $|U_r\rangle$ satisfying
\be \label{eun}
b_0^+ |U_r\rangle =0, \quad b_0^+\wh Q_B |U_r\rangle=0,
\quad \wh Q_B \sum_r a_r |U_r\rangle \ne 0
\, ,
\ee
for any choice of $\{a_r\}$ other than $a_r=0$ for every $r$.
\item Physical states: These are states satisfying
\be \label{eph}
b_0^+ |P_a\rangle=0, \quad \wh Q_B |P_a\rangle =0,
\quad
\sum_a c_a |P_a\rangle \ne \sum_r d_r\,  \wh Q_B|U_r\rangle
\ee
for any choice of $\{c_a\}$, $\{d_r\}$ other than $c_a=0$ for every $a$
and $d_r=0$ for every $r$.
\item Pure gauge states: These are states of the form $\wh Q_B |U_r\rangle$. These are
automatically annihilated  by $\wh Q_B$
due to \refb{enil} and
by $b_0^+$ due to \refb{eun}. 
\end{enumerate}
Note that for 
any given momentum if there is an unphysical state $|U_r\rangle$, there is also a pure
gauge state $\wh Q_B |U_r\rangle$. Generically we expect no other degeneracy
but we shall proceed without making this assumption. A similar classification can be
done for the candidates for $\langle\Psi_m|$.

Let us now 
suppose that at some given momentum at which $\Delta$ has pole,
there are a certain number of
linearly independent 
physical states $\{|P_a\rangle\}$, unphysical states $\{|U_r\rangle\}$ 
and pure gauge states $\{\wh Q_B |U_r\rangle\}$. 
The normalization of these states is chosen arbitrarily.
Then the general form of
$\Delta_0$ is given by
\be \label{eDel}
\Delta_0 = \sum_a |P_a\rangle \langle B_a| 
+ \sum_r |U_r\rangle \langle C_r| + \sum_r \wh Q_B |U_r\rangle \langle D_r|\, ,
\ee
for some states $|B_a\rangle,|C_r\rangle,|D_r\rangle\in\wh\HH_T$.
Eq.\refb{epsbm} now gives
\be \label{eid22}
 \langle\Psi| b_0^+ =0\, , \quad 
\langle\wh Q_B \Psi| b_0^+  = 0  \, , \quad \hbox{for 
  $\langle\Psi|= \langle B_a|$, $\langle C_r|$ or $\langle D_r|$}\, .
\ee
Substituting \refb{eDel} into
\refb{eqdid} and using \refb{eun}, \refb{eph} we also get
\be
\sum_r \wh Q_B |U_r\rangle \langle C_r| c_0^-+
 \sum_a |P_a\rangle \langle B_a| c_0^-  \wt Q_B + 
 \sum_r |U_r\rangle \langle C_r|  c_0^-  \wt Q_B+ \sum_r \wh Q_B |U_r\rangle \langle D_r| 
  c_0^-  \wt Q_B= 0\, .
\ee
Using this, and
the fact that $\{|P_a\rangle\}$, $\{|U_r\rangle\}$ 
and $\{\wh Q_B |U_r\rangle\}$ are linearly independent, we get
\ben \label{erel}
\langle B_a| c_0^-  \wt Q_B =0 \quad &\Rightarrow&
\quad \langle\wh Q_B B_a|=0, 
\nonumber \\  
\langle C_r| c_0^- = - \langle D_r|  c_0^-  \wt Q_B \quad &\Rightarrow&
\quad \langle C_r| =
- (-1)^{{D_r}} \langle \wh Q_B D_r|  , \nonumber \\
\langle C_r| c_0^- \wt Q_B=0  \quad &\Rightarrow&
\quad \langle \wh Q_B C_r|=0\, .
\een
$(-1)^{D_r}$ takes value 1 if $D_r$ is 
grassmann even  and $-1$ if $D_r$ is
grassmann odd.
The last  equation in \refb{erel} 
in fact follows from the second equation and nilpotence of $\wt Q_B$.
Using the second equation in \refb{erel} we can rewrite \refb{eDel} as
\be \label{edel0}
\Delta_0 = \sum_a |P_a\rangle \langle B_a| 
- \sum_r (-1)^{  {D_r}}  |U_r\rangle \langle \wh Q_B D_r|
+ \sum_r \wh Q_B |U_r\rangle \langle D_r|\, .
\ee
Since according to \refb{edefDelta} $\Delta$ carries total ghost number $-2$ and
since the BPZ inner product pairs
states carrying total ghost number 6,  we have
\be \label{enghost}
n_{P_a}+n_{B_a}=4, \quad n_{U_r}+n_{D_r}=3\, ,
\ee
where for any state $|A\rangle$, $n_A$ denotes its ghost number.

The generalization of this analysis
to type II string theories is straightforward\cite{1508.02481}.
The string field will now have 
four sectors satisfying NSNS, NSR, RNS and RR boundary conditions. $\GG$ will 
be given by the identity operator in the NSNS sector, zero mode of the right-handed
PCO in the NSR sector, zero mode of the left-handed PCO in the RNS sector
and the product of the zero modes of the left-handed and right-handed PCO's in the
RR sector. 
$\wh\HH_T$ and $\wt\HH_T$ will be defined as
\be
\wh\HH_T =\HH_{-1,-1}\oplus \HH_{-1,-1/2} \oplus \HH_{-1/2.-1}\oplus \HH_{-1/2, -1/2},
\quad 
\wt\HH_T =\HH_{-1,-1}\oplus \HH_{-1,-3/2} \oplus \HH_{-3/2,-1}\oplus \HH_{-3/2, -3/2},
\ee
where $\HH_{m,n}$ denotes the subspace of $\HH_T$ carrying left-handed picture number $m$
and right-handed picture number $n$.
The rest of the analysis remains unchanged.

\sectiono{Unitarity} \label{sproof}

In this section we shall prove unitarity of the amplitudes of superstring field theory.
This analysis will be valid for both heterotic and type II string theories.

Let $ \Gamma^{(N)}(
| A_1\rangle, \cdots | A_N\rangle) $ denote 
the truncated Green's function in which the external leg propagators are
removed.
$\Gamma^{(N)}$ satisfies the following Ward identity\cite{1508.02481}:
 \be \label{egrid}
\sum_{i=1}^N (-1)^{\gamma_1+\cdots \gamma_{i-1}} \Gamma^{(N)}(
| A_1\rangle, \cdots | A_{i-1}\rangle,
\wh Q_B| A_i\rangle, | A_{i+1}\rangle, \cdots | A_N\rangle) =0\, ,
\ee
where $\gamma_i$ is the grassmannality of $A_i$.

\begin{figure}
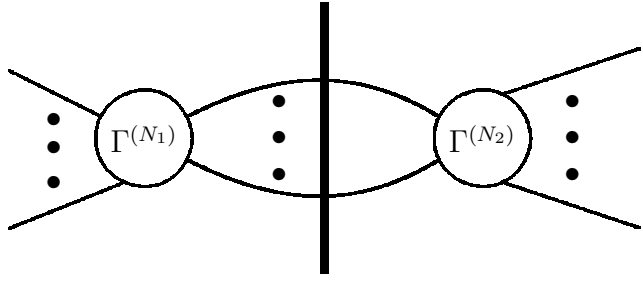

\begin{center}

\figtwo

\end{center}

\vskip -.8in

\caption{A cut diagram in superstring field theory. \label{f2}}

\end{figure}

Now in a cut diagram like the one shown in Fig.~\ref{f2},
each cut propagator is replaced by the right hand side of \refb{edel0}
together with a $2\,\pi\,\delta(k^2+m^2)\, \theta(k^0)$ factor. 
Let us suppose that we have a cut diagram with $N$ cut propagators.
Using the
superscript $(i)$ to label the states associated with the $i$-th cut propagator
and the operators acting on these states, we have a net factor of
\be \label{ebbi}
\prod_{i=1}^N (\Delta_0)^{(i)} =
\prod_{i=1}^N  \left[ \sum_a |P_a^{(i)}\rangle \langle B^{(i)}_a| 
- \sum_r (-1)^{  {D_r^{(i)}}} |U_r^{(i)}\rangle 
\langle \wh Q_B^{(i)} D^{(i)}_r|  \,
+ \sum_r \wh Q_B^{(i)} |U_r^{(i)}\rangle 
\langle D^{(i)}_r|  
 \right]\, ,
\ee
associated with all the cut propagators.\footnote{The range of $a$ and $r$ in \refb{ebbi} 
are in general different for different $i$.}
In this 
the
ket states are inserted into the amplitude $\Gamma^{(N_1)}$ 
on the left side of the cut and the bra states are inserted into
the amplitude $\Gamma^{(N_2)}$ on the right side of the cut. 
Besides these $\Gamma^{(N_1)}$ and $\Gamma^{(N_2)}$ have insertions of
external incoming and outgoing states respectively, which are all annihilated by
$\wh Q_B$.

We now expand \refb{ebbi} as a sum of $3^N$ terms. 
There is one term given by 
\be \label{edesd}
\prod_{i=1}^N \left\{ \sum_a |P_a^{(i)}\rangle \langle B^{(i)}_a| \right\} \, .
\ee
Each of the other terms
has a certain number
(say $K<N$) 
of factors of $ \sum_a |P_a^{(i)}\rangle \langle B^{(i)}_a|  $. 
We group together all terms with
the same factors of $ \sum_a |P_a^{(i)}\rangle \langle B^{(i)}_a|  $, and in any given
group we denote by $S$ the set
of labels $i$ carried by the rest of the factors. $S$ contains $N-K$ elements. 
We separate out from $S$ one particular label which we call $\alpha$. 
For definiteness we can take $\alpha$ to be the lowest element of $S$.
For any $A\subseteq S -\{\alpha\}$, where $S_1-S_2$
for $S_2\subseteq S_1$
denotes the set $S_1$ with the elements in $S_2$ removed,
let us denote
by $F_S(\alpha; A)$ the amplitude associated with the cut diagram
where the label $\alpha$ is carried by the factor 
$ - \sum_r (-1)^{  {D_r^{(\alpha)}}}  |U^{(\alpha)}_r\rangle 
\langle\wh Q_B^{(\alpha)}D^{(\alpha)}_r|  $, 
the labels $i$ in $A$ are carried by $-\sum_r 
(-1)^{D^{(i)}_r}|U_r^{(i)}\rangle \langle  \wh Q_B^{(i)}D^{(i)}_r| $
and the labels $i$ 
in $S-\{\alpha\}-A$ are carried by $\sum_r \wh Q_B^{(i)}|U_r^{(i)}\rangle \langle D^{(i)}_r|$. 
Similarly we denote by $G_S(\alpha; A)$ the amplitude 
where the label $\alpha$ is carried by $\sum_r
\wh Q_B^{(\alpha)}|U^{(\alpha)}_r\rangle 
\langle D^{(\alpha)}_r|  $, 
the labels $i$ in $A$ are carried by $-\sum_r (-1)^{D^{(i)}_r}
|U_r^{(i)}\rangle \langle  \wh Q_B^{(i)} D^{(i)}_r| $
and the labels $i$ in $S-\{\alpha\}-A$ are carried by 
$\sum_r \wh Q_B^{(i)}|U_r^{(i)}\rangle \langle D^{(i)}_r|  $. 
Then the sum of all terms in a given group, 
i.e.\ with a fixed set of labels $i$
carrying $ \sum_a |P_a^{(i)}\rangle \langle B^{(i)}_a|  $ 
factors,
is given by
\be \label{epair}
\sum_{A\subseteq S-\{\alpha\}} \left[ F_S(\alpha; A) + G_S(\alpha; A)\right]  \, .
\ee

Let us consider the amplitude $F_S(\alpha; A)$. In this case
the insertion associated with the
line $\alpha$ to the amplitude $\Gamma^{(N_2)}$
on the right of the cut is 
$-(-1)^{D^{(\alpha)}_r}\langle \wh Q_B^{(\alpha)}D^{(\alpha)}_r| $. The other insertions
involve the states $\langle B^{(i)}_a|  $ for $i\not\in S$, the states 
$-(-1)^{D^{(i)}_r} \langle  \wh Q_B^{(i)} D^{(i)}_r|$ with $i\in A$, the states
$\langle D^{(i)}_r|$ for $i\in S-\{\alpha\}-A$ and the external physical states which are all
annihilated by $\wh Q_B$. Now we can use \refb{egrid} to express this amplitude as a
sum of terms
in which 
$\wh Q_B^{(\alpha)}D^{(\alpha)}_r$ is replaced 
by $D^{(\alpha)}_r$, but $\wh Q_B$ acts in turn on the other states.
Since the external states as well as $B_a^{(i)}$ and $\wh Q_B^{(i)} D^{(i)}$ are
all annihilated by $\wh Q_B$,
the only
non-vanishing contribution comes from the terms where 
$\wh Q_B$ acts on one of the states 
$\langle D^{(j)}_r|  $ for $j\in S-\{\alpha\}-A$. This gives,
\be \label{efid}
F_S(\alpha; A) = \sum_{j\in S-\{\alpha\}-A} \, s(\alpha; j; A) \, H_S(\alpha; j; A)
\ee
where $s(\alpha; j; A)$ takes value $\pm 1$ and
$H_S(\alpha; j; A)$ denotes an amplitude where the label $\alpha$ is carried by
$\sum_r |U^{(\alpha)}_r\rangle 
\langle D^{(\alpha)}_r|$, the label $j$ is carried by
$-\sum_r (-1)^{D^{(j)}_r}
\wh Q_B^{(j)}|U^{(j)}_r\rangle \langle \wh Q_B^{(j)}D^{(j)}_r|$, the labels $i$ in $A$ are
carried by
$-\sum_r (-1)^{D^{(i)}_r} |U_r^{(i)}\rangle \langle \wh Q_B^{(i)}D^{(i)}_r|$
and the labels $i$ in $S-\{\alpha\}-A-\{j\}$ are carried by 
$\sum_r \wh Q_B^{(i)}|U_r^{(i)}\rangle \langle D^{(i)}_r|$. 
Carrying out a similar manipulation of the amplitude 
$\Gamma^{(N_1)}$ on the left of the cut,
we get
\be \label{egid}
G_S(\alpha; A) = \sum_{j\in A} s' (\alpha; j; A-\{j\}) H_S(\alpha; j; A-\{j\})\, .
\ee
where $s' (\alpha; j; A-\{j\})$ takes value $\pm 1$. This gives
\ben \label{eone}
\sum_{A\subseteq S-\{\alpha\}} F_S(\alpha; A) 
&=& \sum_{A\subseteq S-\{\alpha\}} \sum_{j\in S -A-\{\alpha\}} 
s(\alpha; j; A) \, H_S(\alpha; j; A)
\nonumber \\
&=& \sum_{j\in S-\{\alpha\}} \sum_{A\subseteq S-\{\alpha, j\}}
s(\alpha; j; A) \, H_S(\alpha; j; A)\, ,
\een
and
\ben \label{etwo}
\sum_{A\subseteq S-\{\alpha\}} G_S(\alpha; A) 
&=& \sum_{A\subseteq S-\{\alpha\}} \sum_{j\in A} s'(\alpha; j; A-\{j\})\,  H_S(\alpha; j; A-\{j\})
\nonumber \\
&=& \sum_{j\in S-\{\alpha\}} \sum_{A\subseteq S-\{\alpha, j\}} s'(\alpha; j; A)\, 
H_S(\alpha; j; A)\, ,
\een
where in the last step we have relabelled $A-\{j\}$ as $A$.
The right hand sides of
 \refb{eone} and \refb{etwo} are the same up to signs. We shall now show that
the signs are such that these terms cancel pairwise in \refb{epair}. 

The manipulations in \refb{eone} involve the rearrangement
\ben \label{eonea}
&& -(-1)^{D^{(\alpha)}_r} |U^{(\alpha)}_r\rangle 
\langle  \wh Q_B^{(\alpha)}D^{(\alpha)}_r|  
\left\{
\prod_{i; \alpha<i<j} (\Delta_0)^{(i)}  \right\}
\wh Q_B^{(j)}|U^{(j)}_r\rangle \langle D^{(j)}_r|  
\nonumber \\
&\Rightarrow& (-1)^{D^{(j)}_r}|U^{(\alpha)}_r\rangle 
\langle D^{(\alpha)}_r|  \, 
\left\{
\prod_{i; \alpha<i<j} (\Delta_0)^{(i)} \right\}
\wh Q_B^{(j)}|U^{(j)}_r\rangle \langle 
 \wh Q_B^{(j)} D^{(j)}_r|
 \, .
\een
The sign on the right hand side is fixed as follows. First there is a minus sign 
from having to take all but one term from the left to the right hand side of
\refb{egrid}. Second there is a factor
of $(-1)^{D^{(\alpha)}_r}$ from having to take $\wh Q_B$ through $D^{(\alpha)}_r$.
These two together cancel the $-(-1)^{D^{(\alpha)}_r}$ factor on the left.
Moving $\wh Q_B$ through 
$\prod_{i} (\Delta_0)^{(i)}$ does not generate a sign since the latter
operator is grassmann even.\footnote{Note that we are not actually 
commuting $\wh Q_B$
through the operators $\Delta^{(i)}$, -- this would generate additional terms
due to \refb{eboth}. 
The transfer of $\wh Q_B$ from one state to another takes place
through the amplitude $\Gamma^{(N_2)}$. However the extra sign picked up due to
the grassmannality of the operators can be determined just from the relative position
of the operators in an expression, and that is the way we are determining the sign.}
Finally moving $\wh Q_B$ through $\wh Q_B^{(j)}|U_r^{(j)}\rangle$
generates a factor of $-(-1)^{U^{(j)}_r}=(-1)^{D^{(j)}_r}$ using \refb{enghost}.. 
This is the factor we see on the right hand side
of \refb{eonea}.

On the other hand manipulations in \refb{etwo} involve the rearrangement
\ben \label{etwoa}
&& -(-1)^{D^{(j)}_r} \wh Q_B^{(\alpha)}|U^{(\alpha)}_r\rangle 
\langle D^{(\alpha)}_r|
\left\{\prod_{i; \alpha<i<j} (\Delta_0)^{(i)} \right\} |U^{(j)}_r\rangle \langle   
\wh Q_B^{(j)} D^{(j)}_r| 
\nonumber \\
&\Rightarrow& -(-1)^{D^{(j)}_r} |U^{(\alpha)}_r\rangle 
\langle D^{(\alpha)}_r|  
\left\{\prod_{i; \alpha<i<j} (\Delta_0)^{(i)}  \right\}
\wh Q_B^{(j)}|U^{(j)}_r\rangle \langle   \wh Q_B^{(j)}D^{(j)}_r| 
\, .
\een
In this manipulation two minus signs cancel. 
First of all we get a minus sign from having to take all but one term in
\refb{egrid} from the left to the right side. Since $|U^{(\alpha)}_r\rangle 
\langle D^{(\alpha)}_r|$ 
is a grassmann odd operator due to \refb{enghost}, passing $\wh Q_B$ through this
generates a second minus sign. Therefore the right hand side of \refb{etwoa} has
the same sign as the left hand side.

We now see that the the right hand sides of \refb{eonea} and
\refb{etwoa} cancel. This cancelation works for every term in \refb{eone} and \refb{etwo},
making \refb{epair} vanish. This
shows that the only term that contributes is the one where \refb{ebbi} is replaced by \refb{edesd}.

This still does not prove that only physical states contribute since the only information about 
$\langle B_a|$ that we have is from \refb{eid22} and
the first equation in \refb{erel}, and this allows 
$\langle B_a|$ to be either a physical state or
a pure gauge state of the form $\langle \wh 
Q_B E_a|$ for some $\langle E_a|$.
However since all other states entering in the argument of
$\Gamma^{(N_2)}$ are annihilated by $\wh Q_B$, the 
amplitude with one or more $\langle B_a|$
having the form $\langle\wh Q_B E_a|$ will
vanish due to \refb{egrid}. This shows that $\langle B_a|$ must be a physical state.
It now follows from \refb{edesd} that only physical states contribute to the cut propagators.
This is the desired result that establishes unitarity of the amplitude.

\sectiono{Discussions} \label{sdiss}

In this paper we have completed the proof of unitarity of covariant superstring 
field theory. Therefore this theory represents a manifestly Lorentz invariant, 
ultraviolet finite and unitary theory. Furthermore infrared divergences associated 
with tadpoles and mass renormalization can be dealt with using standard quantum
field theory techniques.

We must note however that 
when the number of non-compact space-time dimensions
$D$ is 4 or less, the S-matrix suffers from the usual infrared divergences
and we have to carry out the usual procedure of summing over final states and averaging
over initial states to get a finite result for physical cross 
section\cite{kinoshita,lee,bloch,sterman}.  This has not yet 
been worked out in superstring field
theory. We hope to return to this problem in the future.

\bigskip

\noindent {\bf Acknowledgement:}
I wish to thank Roji Pius
for useful discussions.
This work was
supported in part by the 
DAE project 12-R\&D-HRI-5.02-0303 and J. C. Bose fellowship of 
the Department of Science and Technology, India.

\end{document}